\documentclass[aps,physrev,preprint,superscriptaddress, nofootinbib, showkeys]{revtex4-2}

\usepackage{graphicx}

\usepackage{hyperref}

\usepackage{etoolbox}
\AtBeginEnvironment{quote}{\par\small}

\begin{document}


\title{Making sense of quantum teleportation: An intervention study on students' conceptions using a diagrammatic approach}


\author{Sebastian Kilde-Westberg}
\email{sebastian.kilde.westberg@physics.gu.se}
\author{Andreas Johansson}
\affiliation{Department of Physics, University of Gothenburg, SE 412 96 Gothenburg, Sweden}
\author{Anna Pearson}
\affiliation{Quantinuum, Oxford, UK}
\author{Jonas Enger}
\affiliation{Department of Physics, University of Gothenburg, SE 412 96 Gothenburg, Sweden}



\begin{abstract}
\noindent
Quantum physics education at the upper-secondary level traditionally follows a historical approach, rarely extending beyond early 20th-century ideas, leaving students unprepared for comprehending modern quantum technologies central to everyday life and many facets of modern industry.
To address this gap, we investigated how upper-secondary students and pre-service teachers understand quantum teleportation when taught with a simplified diagrammatic formalism based on the ZX-calculus, which represents quantum processes as diagrams of wires and boxes.
Through phenomenographic analysis of video-recorded group work sessions, written responses to exercises, and a group interview, with a total of $n=21$ participants, we identified an outcome space consisting of four qualitatively different, hierarchically ordered categories of description encapsulating the different ways of experiencing quantum teleportation.
The categories revealed that a conceptual progression depends on how one understands the temporality in quantum processes, the role of entanglement in quantum teleportation, the active nature of quantum measurements, and interpretations of mathematical operations in the diagrams.
Our findings demonstrate that while a simplified diagrammatic formalism for teaching quantum physics provides an accessible entry point at the upper-secondary level, it does not automatically resolve fundamental conceptual challenges, and requires careful consideration in terms of developing teaching and learning sequences.
Finally, these results provide educators with a deeper understanding of conceptual affordances and challenges for designing and improving instruction, whilst also highlight the need for further exploring how students and teachers alike understand quantum phenomena.
\end{abstract}

\keywords{Quantum education research, phenomenography, quantum teleportation, upper-secondary school physics, pre-service teachers, conceptual understanding, diagrammatic formalism}

\maketitle

\section{Introduction}
\label{sec:introduction}
\noindent
Today, quantum physics (QP) lies at the heart of many technological developments, from material science to computers and lasers \cite{Jaeger2018}.
This has led to an increased need for industry and educational institutions to educate the future quantum workforce in areas ranging from having a solid foundation of quantum concepts, to more practical knowledge about working with hardware and software applications relying on quantum technologies \cite{Chen2023, Greinert2023}.
Since quantum technologies rely heavily on physics, there has been a growing need within the physics education research (PER) community to develop and explore new approaches to teaching various aspects of QP, including how to better introduce quantum concepts pre-university \cite{Lane2025}.

In Europe, QP is commonly introduced in upper-secondary schools, predominantly through a historical or a semi-historical path that follows the footsteps of the theory development in the early 1900s \cite{Stadermann2019, Michelini2021}.
Following such an approach, it is rare for students to move much beyond the ideas of the Bohr atomic model with discretized energy levels being thought of as synonymous to the solar system model.
Further, a recent study highlights, from a multi-stakeholder perspective including QP experts, PER researchers, and upper-secondary school teachers, that concepts such as superposition, qubits, entanglement, and quantum measurement should be among the key foci when teaching QP at the upper-secondary school level, whereas mathematical foundations are regarded as less important \cite{Merzel2024}.
The same study also highlights that upper-secondary school teachers seem to have little knowledge of illustrative examples of superposition, that can be created in a Mach-Zehnder interferometer or with a qubit, and that the teachers participating in the study generally had a rather limited tool set of illustrations and examples of quantum concepts available to them.
Based on these findings, there is a clear need to develop and further explore teaching approaches and illustrative examples that go beyond current curricula and upper-secondary textbooks, as a step toward improving how QP is taught pre-university.
Teaching foundations of quantum computing using a simplified diagrammatic formalism (QPic \footnote{which has been used previously to refer to `quantum picturalism' \cite{Dndar-Coecke2025}.}) adapted from ZX-calculus \cite{Coecke2023} may represent one promising approach.
ZX-calculus is a graphical language for quantum theory that represents quantum processes as diagrams of wires and boxes \cite{Coecke2017}.
Unlike traditional approaches to the mathematical foundations of QP, QPic offers potential pedagogical advantages \cite{Dndar-Coecke2023, Dndar-Coecke2025} by avoiding the complexities of matrices and a level of calculus beyond what is taught at the upper-secondary school level, whilst preserving a level of formal mathematical structure through diagrammatic representations.

The present study investigates whether QPic can address these identified needs by making modern quantum concepts like quantum teleportation---which involves several key concepts, among them entanglement and measurement---accessible to upper-secondary students.
Specifically, we seek to contribute to the current body of research in the field of quantum education with more qualitative insight into how students are experiencing and interpreting teaching approaches with simple mathematical structures.
By further developing and testing proposed teaching materials to introduce pre-university students to modern QP, our study can serve multiple purposes.
First, it provides teachers with practical knowledge about different ways students may conceive quantum concepts and their relation to QPic introduced in \cite{Coecke2023}.
Second, it provides quantum education researchers with deeper qualitative knowledge on ways of understanding quantum phenomena when being taught using varying mathematical formalisms.
Third, it opens up new avenues for educators and researchers alike regarding designing and implementing a diagrammatic approach to teaching QP.

\subsection{Teaching quantum physics beyond the historical approach: a review of needs and proposals}
\noindent
In this paper, we define current approaches to teaching QP as ones following what is typically found in textbooks, which is the historical or semi-historical approach introduced above \cite{Stadermann2019, Michelini2021}.
Here, we provide a review of identified needs and recent investigated approaches to introducing QP concepts in primarily upper-secondary school physics which diverge from current approaches.

In a review on approaches to teaching and learning QP as discussed in the PER literature \cite{Michelini2021} three major approaches were identified: a \textit{historical approach}, a \textit{formal structural approach}, and a \textit{conceptual approach}.
The \textit{formal structural approach} generally follows a similar path as what is taught at university-level QP courses with heavy emphasis on mathematical structure, a focus on the wave function and its predictive power of quantum phenomena.
It is an approach that has similarities to the \textit{historical approach}, but puts more clear emphasis on discussing QP using a modern view of matter and radiation and requires students to be well versed in the mathematical skills of manipulating the wave function.
On the other end of the spectrum we find the \textit{conceptual approach}, which shifts the focus towards phenomenological explanations of various experiments manipulating two-state systems, such as the Stern-Gerlach experiment, the Mach-Zehnder interferometer, and the double-slit experiment.
The aim of this approach is typically to promote a new way of thinking about the physical world, grounded in the idea that conceptual assumptions students make use of in classical physics are different, albeit sometimes lexically similar, from what is required to successfully comprehend quantum phenomena.
In a bibliometric analysis giving an overview of the publication landscape and scope in QP education research, we find that research aligned with the \textit{conceptual approach} seems most promising for tackling a prevalent problem in the era of modern quantum technologies \cite{Bitzenbauer2021}.
Namely, that today there is a growing need for a basic understanding of quantum technologies, which requires a degree of conceptual understanding of quantum concepts but less emphasis on mathematical foundations, for instance, to enable individuals to work with and program quantum computers.

To better understand educational needs at the upper-secondary level, the European  Competence Framework for Quantum Technologies provides useful context \cite{EuropeanCommisisionDirectorate-GeneralforCommunicationsNetworks2024}, which has similarities to the U.S. quantum workforce development strategy \cite{QIST2022}.
The  framework identifies that different career paths in the quantum industry require  diverse combinations of expertise—from developing theoretical foundations and building quantum computer components to designing quantum software or working in business and strategic roles. 
This highlights the need for QP education to prepare students by fostering an awareness of diverse career opportunities in quantum technologies.
The diversity in skills required by the quantum industry is further evidenced by a recent study of quantum industry demands by analyzing job postings over six months of 37 quantum technology industry companies \cite{Devendrababu2025}.
Results from the study indicate that companies seek candidates with a broad set of skills, including strong foundations in QP, applied mathematics, programming skills, as well as experience working with quantum experiments and simulations.
Thus, apart from having a pure physics degree, companies also seek people with some knowledge of quantum foundations, but who have expertise in the field of computer science or various engineering disciplines.

When teaching QP, at the upper-secondary school level and beyond, previous research highlights that students have several difficulties when trying to make sense of fundamental concepts, including wave-particle duality, wave functions, atomic models, and complex behaviors such as time dependence and nondeterminism \cite{Krijtenburg-Lewerissa2017}.
Specifically, student difficulties seem to be closely tied to the challenge of making assumptions and developing argumentation based on quantum rather than classical or semi-classical physics.
To address these conceptual difficulties, several proposals aligned with the \textit{conceptual approach} have been investigated recently targeting the upper-secondary level, including ones such as inquiry-based learning environments for QP \cite{VilartaRodriguez2020}; activities linking classical and quantum computing \cite{Sun2024}, the introduction of a simplified Dirac notation to analyze single-photon interference in a Michelson interferometer \cite{Hennig2024a, Hennig2024b}; game-based approaches to teaching quantum concepts \cite{Marckwordt2021, Lpez-Incera2019, Goff2006, Piispanen2025}; instruction on the basics of quantum cryptography \cite{Weissman2024}; and the use of experiments, simulations, and multiple representations in QP \cite{Marshman2024, Borish2024, Passante2024, Dndar-Coecke2023}.
What these proposals have in common is adhering to the notion that it is challenging, yet crucial, for conceptual approaches to have connections to mathematical formalism, regardless if it is explicated in the teaching or not.
Further, there seems to be agreement that to address the many conceptual hurdles in QP, students need exposure to experiments, conceptual arguments, mathematical rigor, and various representations that highlight the uses and limitations of models and analogies.
However, currently lacking from proposals on alternative teaching approaches to QP is a clear connection to how such proposals affect students' general view and understanding of the concepts and phenomena in focus.
Rather, findings and conclusions connect to addressing specific previously identified conceptual hurdles, or misconceptions, but give less insight into a qualitative understanding of what the participating students' views of the phenomena under study are after an intervention, and how the intervention aligns with conceptual views of experts.

\subsection{Aim and research question}
\noindent
The current study investigates how QPic can be used to teach introductory QP. 
Specifically, we aim to gain essential qualitative insights regarding ways of experiencing quantum teleportation in the context of it being taught using QPic similar to Chapter 1 in Quantum in Pictures \cite{Coecke2023}.
Such insights can then be used to design further educational material, and serve as an important starting point for teachers about understanding how their students may understand and interpret the material being taught.
We approach this aim by exploring the following research questions:

RQ1: What are the qualitatively different ways of experiencing a diagrammatic representation of quantum teleportation among upper-secondary students and pre-service physics teachers?

RQ2: In what ways do students relate the use of QPic to their understanding of the principle of quantum teleportation?

\subsection{Theoretical framework}
\label{sec:framework}
\noindent
To answer the research questions, we apply the methodological approach of phenomenography as a way of gaining insight into students' ways of comprehending quantum teleportation, in the context of how the intervention was designed.
Relying on phenomenography in a qualitative study often involves developing \textit{categories of description}, which are to be understood as a limited set of qualitatively different, hierarchically ordered, set of conceptions of a certain aspect of reality \cite{Marton1981}.
These categories collectively form an outcome space that represents the qualitatively different ways in which the phenomenon under investigation is experienced.
In his original paper, Marton \cite{Marton1981} argues that the categories of description can be understood as that a conception is akin to frozen forms of thought, that they exist only in terms of some mental act expressed by someone doing something in a certain setting, or context.
As such, there exists an intimate relationship between someone's conception and their current worldly context, which in phenomenography is studied using a second-order perspective \cite{Marton1997}.
Having a second-order perspective when analyzing data means that one has to consider specific statements as reflecting some specific situation in the world and having to judge it in light of other statements about the same situation.

A phenomenon is commonly understood as being constituted by the `\textit{different ways in which an aspect of the world is conceived or experienced}' \cite{Bruce1999}.
In the phenomenographic tradition, it is considered impossible to experience a phenomenon in the absence of a situation \cite{Marton1997}.
This means that ontologically, phenomenography adheres to a nondualistic view of awareness, wherein people can only describe the world as they experience it.
An implication here is that a phenomenon has to be understood as being viewed through the study participants' awareness and reflections.
It is typical for phenomenographic studies to employ qualitative data collection methods through interviews, open-ended questionnaires, or observations to investigate a phenomenon, that is, how some situation in the world is experienced \cite{Guisasola2023}.
In Sec. \ref{subsec:analysis}, we provide further specific context on how we operationalize phenomenography in the analysis to identify the outcome space presented in the results (Sec. \ref{sec:results}).

\section{An approach to teaching quantum teleportation using diagrams}
\label{sec:context}
\noindent
In this section, we briefly introduce how the intervention used in the current study was designed based on the contents covered in Quantum in Pictures \cite{Coecke2023}.
Specifically, here we aim to give the reader a general understanding of two exercises (E1 and E2) that students worked with, that ended up being the focus of the phenomenographic analysis presented in this paper.
Here, we first cover how the concepts of wires and boxes were introduced, followed by necessary diagrammatic rules that were taught to students in order to provide them with sufficient tools to connect QPic to quantum teleportation.
Then, E1 and E2, that were part of the problem set students worked on during the intervention, are described as they are relevant for understanding the analysis presented in Sec. \ref{sec:results}.
For a more in-depth introduction to the formalism, we refer readers to Appendix A in Dündar-Coecke et al. \cite{Dndar-Coecke2025}, Quantum in Pictures \cite{Coecke2023}, and Picturing Quantum Processes \cite{Coecke2017}.

\subsection{Building blocks and relevant rules}
\label{sec:buildingblocks}
\begin{figure}[b]
\includegraphics[scale=0.25]{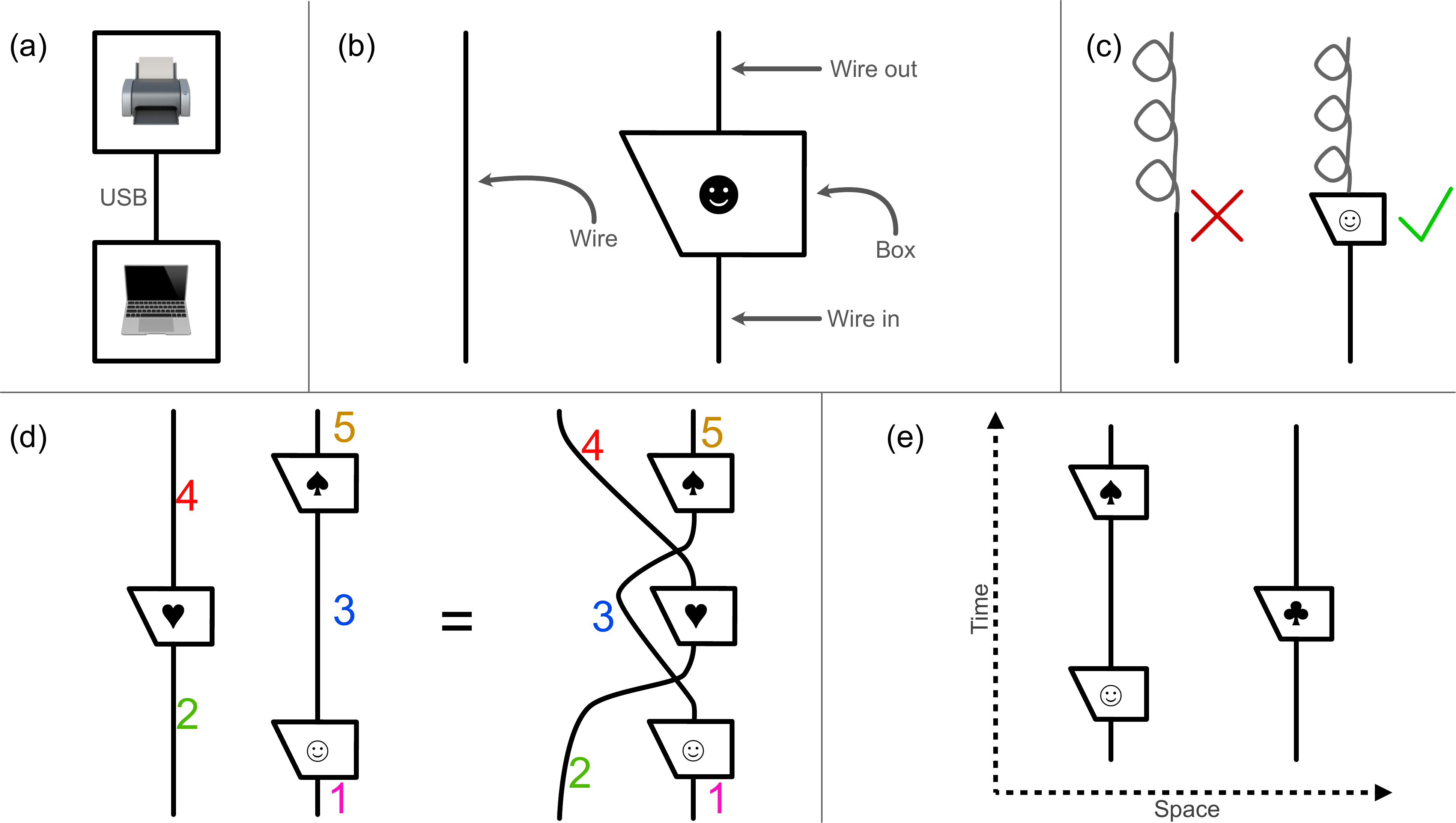}%
\caption{\label{fig:formalism}An overview of some concepts and rules introduced prior to having students use QPic to discuss quantum teleportation. (a) shows how diagrams can be used to convey connections between objects or describe processes in the real world. (b) introduces the two building blocks used in the version of the formalism relevant in the current study, wires and boxes. A wire shows where something is transported, and boxes show what operation is performed on what is transported. (c) establishes the first fundamental rule: type of wires must match, but they can be connected through a box. (d) establishes the second fundamental rule: only connections matter. (e) shows how to conceptualize time and space in diagrams. Parts of the figure, specifically (a), (b), (d), and (e) are adapted from \cite{Coecke2023}.}
\end{figure}

\noindent
In order for any mathematical formalism to be useful, including QPic and the way it is used in the current study, some fundamental rules need to be established, along with necessary building blocks that these rules can be applied to.
These were introduced to the students from a purely mathematical standpoint first, before establishing connections to QP.
First, the concept of diagrams is introduced by discussing how various processes in the world can be visualized using diagrams, such as how to send information from a computer to a printer (Fig. \ref{fig:formalism}a).

Then, the building blocks used in the formalism can be introduced, which are the concepts of wires and boxes (Fig. \ref{fig:formalism}b).
Wires are defined as connections that dictate where something travels, whereas boxes are nodes between connections, indicating that some operation is being done on what is traveling through the wire.
To make diagrams more useful, it is possible to have multiple types of wires, e.g., one signifying water flow, and another electricity, or classical and quantum information.
As a first fundamental rule in the formalism, different types of wires cannot be connected directly, but need some box in between (Fig. \ref{fig:formalism}c).
Further, when constructing diagrams using the formalism, we only need to consider relative connections, which is the second fundamental rule (Fig. \ref{fig:formalism}d).
As a final rule on how to interpret diagrams, the notion of time and space is introduced as the direction of time flows upward, whereas spatial separation between various wires and processes can be communicated by separating them horizontally (Fig. \ref{fig:formalism}e).

\begin{figure}[t]
\includegraphics{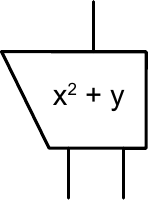}%
\caption{\label{fig:math_example}Establishing a similarity between boxes and mathematical functions. The box represents the function $f(x, y) = x^2 + y$, and produces one output (wire), given two input (wires).}
\end{figure}

\begin{figure}[b]
\includegraphics[scale=0.36]{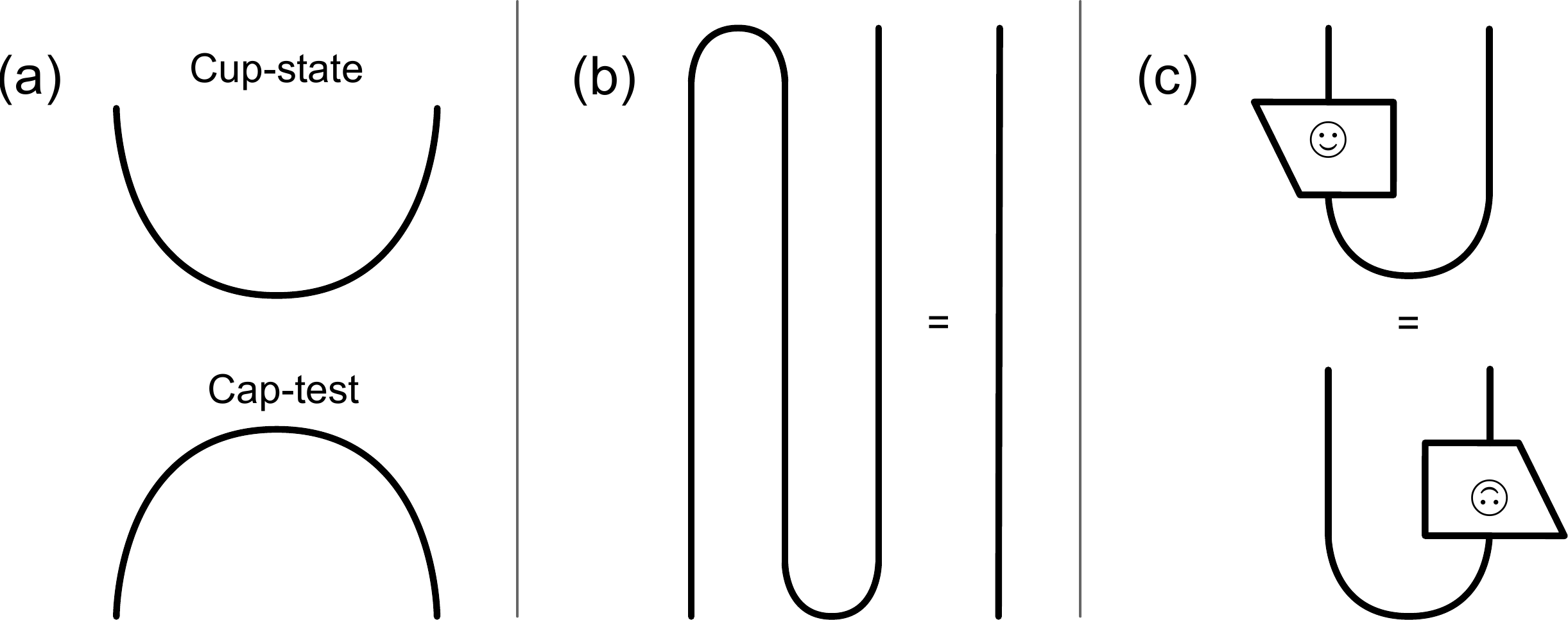}%
\caption{\label{fig:special_boxes}Introducing the ideas of bending wires and sliding boxes. (a) defines a cup-state as a bent wire with two outputs, and a cap-test as a bent wire with two inputs. (b) defines the `yanking equation' which follows logically from the second fundamental rule. (c) illustrates that sliding a box through a cup-state results in a rotation by $\pi$ radians. The figure has been adapted from \cite{Coecke2023}.}
\end{figure}

To illustrate the similarity between boxes and mathematical functions, consider the function $f(x, y) = x^2 + y$.
There, $f(x,y)$ tells us that this function produces some output but needs two inputs, $x$ and $y$.
As such, it can be described using diagrams in terms of a box with two inputs and one output (Fig. \ref{fig:math_example}).
From the example, it is evident that boxes can have various forms regarding connections in and out from them. Boxes with only an output wire are defined as state-boxes, and boxes with only an input wire are defined as test-boxes.
Delving further into boxes, it is also necessary to introduce two more special boxes, namely a cup-state and a cap-test, along with the `yanking equation' \footnote{In \cite{Coecke2017}, Eq. 4.11, there are three `yanking equations.' Here, we only include, and use, the first one.}, and the realization that sliding a box through a cup or cap results in the box being rotated $\pi$ radians (Fig. \ref{fig:special_boxes}).

\subsection{Connecting to quantum teleportation}
\label{sec:connecting_teleportation}
\begin{figure}[b]
\includegraphics[scale=0.43]{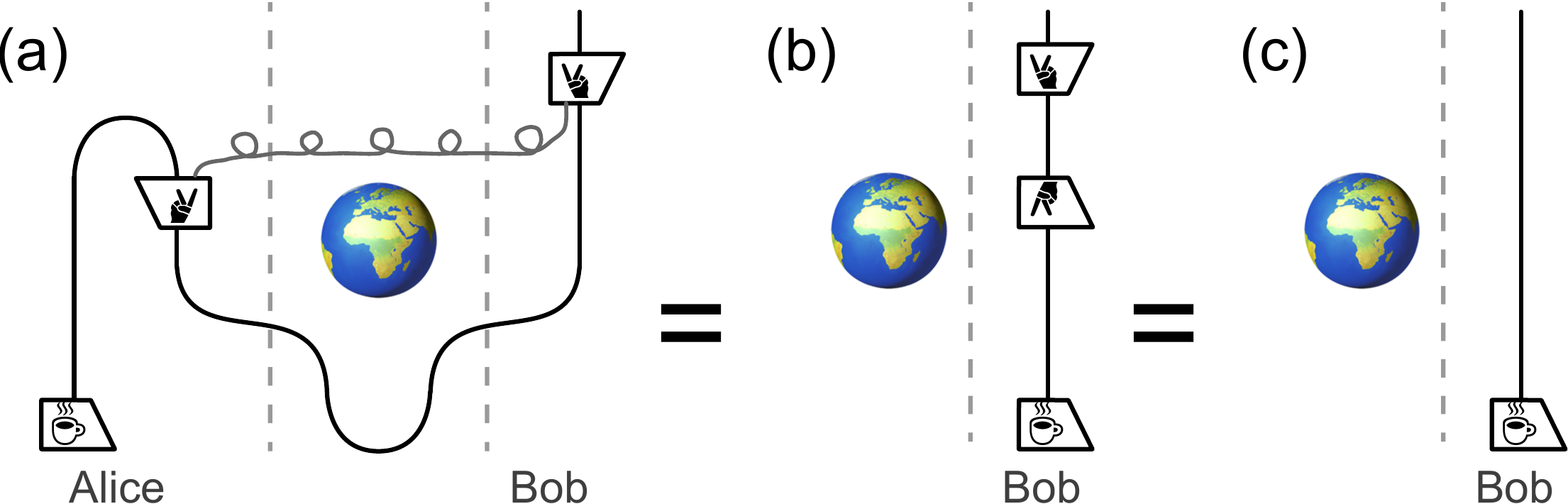}%
\caption{\label{fig:teleportation}A simple quantum teleportation protocol. An emoji of the Earth is included to signify that the three laboratory locations, separated by dashed gray lines, are separated by a significant distance. In (a), the entire process of teleporting a quantum state (state box with a coffee cup symbol) from Alice to Bob is illustrated, with the inclusion of an `error' box (a right hand making the peace sign) being introduced in the BSM, and Bob applying the corresponding `correction' box (a vertically flipped version of the `error' box, i.e., a left hand making the peace sign) after having received classical information about which Bell state Alice ended up with. In (b), we have the mathematical equivalent system after having applied the `yanking equation', and (c) shows that Bob ends up with the original state Alice wanted to teleport since his `correction' box projects his qubit into $|B_{1}\rangle$. The figure has been adapted from \cite{Coecke2023}.}
\end{figure}

\noindent
By employing the building blocks introduced in Sec. \ref{sec:buildingblocks}, it becomes possible to connect the mathematical formalism with QP to illustrate the principle of quantum teleportation (Fig. \ref{fig:teleportation}).
First, the cup-state can be understood as producing entangled pairs of particles.
Then, if applying a box that performs an operation, e.g., measuring polarization direction of photons, measures one photon to be vertically polarized in some axis, the other photon must be horizontally polarized if measured along the same axis, assuming the photons are anti-correlated.
In QPic, this is akin to the fact that when a box moves through a cup-state it is rotated (Fig. \ref{fig:special_boxes}c) and when conducting the same measurement on the other particle, the result is the same as just described, because we have now rotated the `up' direction by $\pi$ radians.

Further, the cap-test represents some kind of measurement, which we here consider to be a Bell state measurement (BSM), that can produce four different outcomes depending on which of the four Bell states is projected.
In QPic, this is represented as cap-tests being `maybe' boxes, which can result in $4$ different outcomes, and thus have a chance of an `error' box being introduced.
Moving on to teleportation, this can be realized using three particles as follows.
Consider two maximally entangled particles that are sent to two different laboratories, $A$ to Alice's and $B$ to Bob's.
The pair of entangled particles are prepared such that they are in the state
\begin{equation}
    \label{eq:entangled_state}
        |\Psi^-_{AB}\rangle = \frac{1}{\sqrt{2}} \left( |0_{A}\rangle |1_{B}\rangle - |1_{A}\rangle |0_{B}\rangle \right) \text{ ,}
\end{equation}
which tells us that if Alice were to do a measurement resulting in $|0\rangle_{A}$, then Bob would, using the same measurement basis, get $|1\rangle_{B}$.
Now, Alice has access to another particle, $A'$ which they want to send to Bob, that is in some state
\begin{equation}
    \label{eq:superposition}
        |\psi\rangle_{A'} = a|0\rangle + b |1\rangle \text{ ,}
\end{equation}
that is a superposition of the states $|0\rangle$ and $|1\rangle$, each with a probability of $|a|^2$ and $|b|^2$ to be found as the result of some measurement \cite{Bouwmeester1997}.
If Alice conducts a BSM on $A'$ and $A$, they are projected onto one of four Bell states,
\begin{equation}
    \label{eq:bell_states}
        |\Psi^{\pm}_{A'A}\rangle\text{, } |\Phi^{\pm}_{A'A}\rangle\text{ },
\end{equation}
each with a probability of $0.25$.
The resulting Bell state is known to Alice, but from this it is impossible to know something about the original state of $A'$.
Further, prior to Alice's measurement, the full state of the three involved particles can be described by
\begin{equation}
    \label{eq:before_measurement}
        |\Psi_{A'AB}\rangle = \frac{1}{2} \left[|\Psi^{-}_{A'A}\rangle |B_1\rangle + |\Psi^{+}_{A'A}\rangle |B_2\rangle + |\Phi^{-}_{A'A}\rangle |B_3\rangle + |\Phi^{+}_{A'A}\rangle |B_4\rangle\right] \text{ },
\end{equation}
where $|\Phi^{\pm}_{A'A}\rangle$ and $|\Psi^{\pm}_{A'A}\rangle$ are the Bell states from Eq. \ref{eq:bell_states}, and $|B_i\rangle \text{, } i=1, 2, 3, 4$ the states Bob's particle $B$ could be in \cite{Bennett1993}.
Following this, when Alice conducts a BSM the outcome is a projection onto one of the four Bell states, it is given by Eq. \ref{eq:before_measurement} that if Alice communicates their results from the BSM to Bob (which can be done via classical communication channels), Bob instantly knows which of the four possible states $B$ have been projected onto.
Bob can then manipulate $B$ by rotation to effectively transform the state $|B_i\rangle$ into $|B_1\rangle$, which is equal to the original state of $A'$ given the original entanglement of $A$ and $B$ was prepared such that they correspond to Eq. \ref{eq:entangled_state}.

Connecting to QPic, depending on information sent to Bob from Alice there are three possible rotational transformation Bob may need to perform, and one outcome in which Bob had to do nothing in order to get the teleported state $|B_1\rangle$.
As such, the quantum teleportation protocol described here can be constructed diagrammatically as seen in Fig. \ref{fig:teleportation}a, with one of three different `error' boxes or no box being present as part of the `cap' box.

\subsection{Student exercises}
\noindent
During the intervention in this study, students had to work through problem sets.
Here, we describe E1 and E2, that are closely related to Fig. \ref{fig:teleportation}.

\begin{figure}[b]
\includegraphics[scale=0.6]{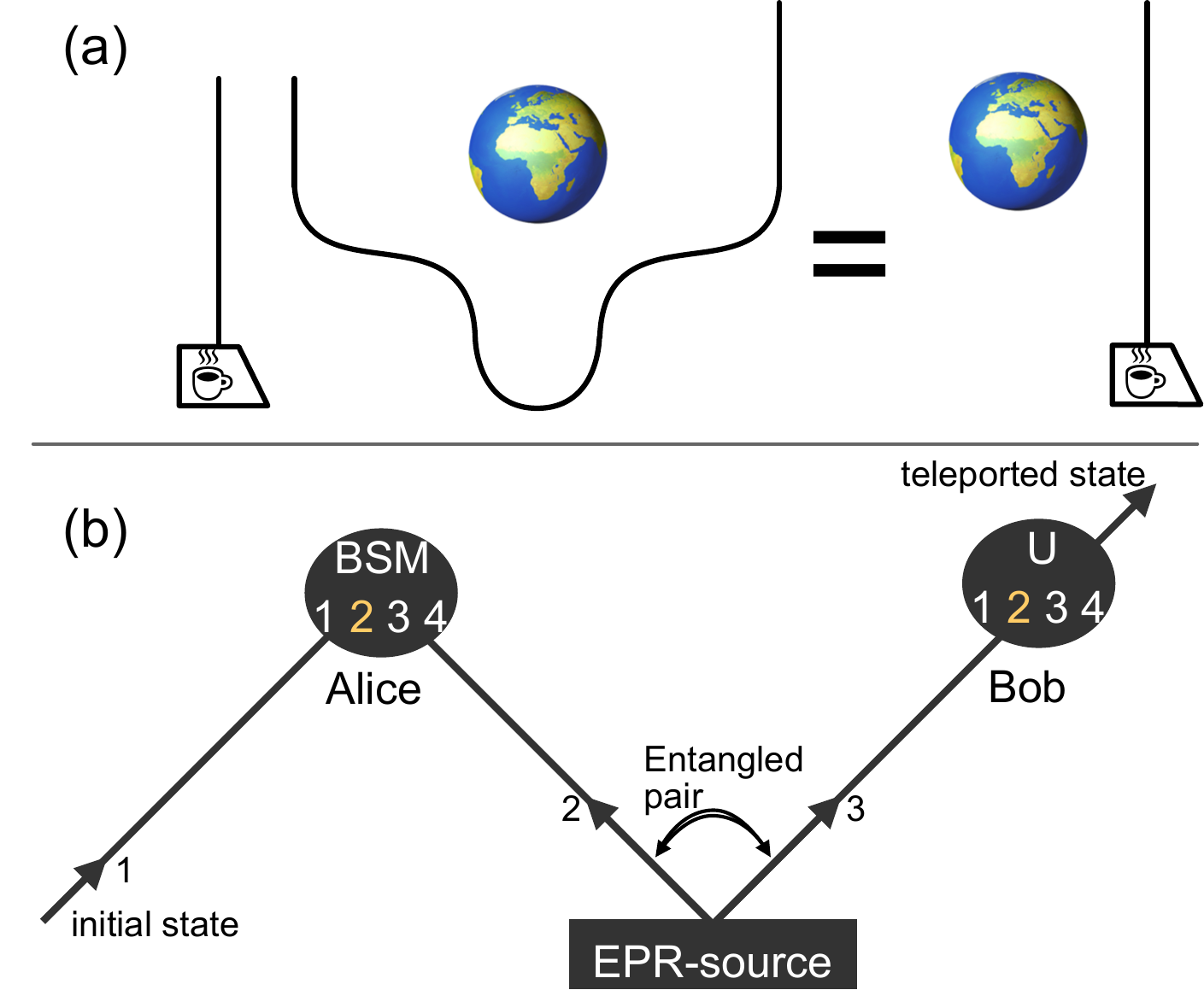}%
\caption{\label{fig:exercises}Adaptions of the figures in E1 (a) and E2 (b). (a) is adapted from \cite{Coecke2023}, and (b) is adapted from \cite{Bouwmeester1997}.}
\end{figure}

In E1, students are provided with Fig. \ref{fig:exercises}a.
The exercise consists of two parts, where the first asks what needs to be done with the diagram (Fig. \ref{fig:exercises}a) in order for the equality to be true.
Part 2 then asks the following: If you know that the diagram describe some process involving quantum particles, do you need to change anything in your proposed solution to part 1?
This requires the solution to include the possibility of `error' and `correction boxes,' as well as a classical communication wire between them.

E2 is a three part exercise, where the first asks students to describe what they think happens in the experiment that is described by Fig. \ref{fig:exercises}b.
Part 2 asks students to construct a diagram using wires and boxes that describes the same experimental situation, essentially having them reconstruct the figure produced in E1.
Finally, part 3 asks if teleportation happens faster or slower than the speed of light.

\section{Method}
\label{sec:method}
\noindent
The study was designed as a two-lesson intervention study, with the explicit aim to introduce students to quantum teleportation.
Lesson one was divided up into two parts; a 30 minute lecture followed by a 15 minute session where students had time to work on a first problem set.
Lesson two had a similar division, but the initial lecture here was instead 15 minutes followed by 20 minutes of group work with a second problem they had not seen previously, which included E1 and E2 as described above.
The two lessons were separated by one week, and all students were encouraged to study the Power Point slides, chapters 1 and 2 in the textbook \cite{Coecke2023}, keep working on the first problem set between lessons one and two, and contact the lecturer (the first author).
During the group work section of each lesson, students were free to ask the teachers questions (the first and last authors). However, during lesson two, the students were told they would not be provided with complete answers to exercises, nor indication if their answers were correct.
Finally, as a way for the researchers to gauge the level of difficulty of the intervention, a pre- and post-test was distributed at the beginning of lesson one, and at the end of lesson two, with five questions about quantum measurements, and one self evaluation question asking students to state the difficulty of the two lessons (post-test only).
The test served as a tool to certify that the material covered during the lesson was not unexpectedly difficult (the mean difficulty rating was $6.3$ out of $10$), and that students had either some or no prior knowledge about the content that was taught.

\subsection{Participants, data collection, and ethical considerations}
\noindent
We invited two different groups of to participate to ensure a breadth of prior knowledge: third-year upper-secondary school students ($n=18$) and fourth-year pre-service teachers ($n=3$).
The upper-secondary school students were part of a group of students taking an elective advanced physics course.
This meant that we could expect that students had been introduced to some QP concepts, mainly stemming from ideas from early QP such as energy quantization in the Bohr atomic model, but also concepts such as superposition, the Heisenberg uncertainty principle,  wave-particle duality, tunneling, and even the Schrödinger wave equation.
Regarding the fourth-year pre-service teachers, they were students who had recently passed a course on modern physics and an introductory quantum mechanics course with a wave-first approach following the structure of Griffiths \cite{Griffiths2018}.
They were invited to ensure we had participants with an even deeper prior knowledge about QP.

The current study draws upon data collected during lesson two, and answers to an exam question given at the end of the upper-secondary school students course.
During lesson two, we collected students' hand written answers and notes during the group work part of the lesson.
For the exam questions, copies of pseudonynomized answers of participating students to one exam question were collected.
Further, the group work session was also video recorded using six cameras, one placed above each group's workbench.
Finally, the pre-service teachers were also invited to partake in a video recorded group interview shortly after concluding lesson two.
The interview served as an opportunity for the first author to gain additional insight into the students understanding of the phenomenon under study.

In designing the current study, we adhered to local and national ethical guidelines \cite{SwedishResearchCouncil2025}.
All participants were given verbal and written information about the study and the type of data to be collected, and invited to ask questions before deciding if they wanted to participate.
Students who agreed to partake in the study all gave their written consent, and could rescind their consent at any time.

\subsection{Analysis and limitations}
\label{subsec:analysis}
\noindent
The aim of the current study is to gain a deeper understanding of students' understanding of quantum teleportation.
Relying on phenomenography as our approach to gain such insight, a central goal of the analysis is to develop categories of description.
Practically, this was done by one of the researchers (the first author) first by going through all data collected in order to familiarize themselves with the content and identify initial underpinnings of the outcome space.
During this initial part, the goal was to initialize the identification of a pool of meaning, which is a way of identifying connections between specific instances of students expressions episodes of discussion or problem solving that helps identify a specific way of understanding, or communicating about, the phenomenon under study \cite{Adawi2006, Holmqvist2019}.
Since the pool of meaning is comprised out of a collection of specific episodes or expressions, the same student may, within the data collected, express many different ways of understanding the phenomenon.
As such, in phenomenographic research, it is essential to understand that categories of description are not a way to categorize the students as a whole; rather, it is to uncover the variation in ways of conceptualizing the phenomenon \cite{Adawi2006, Guisasola2023}.

After having identified a pool of meaning, a preliminary set of six categories of descriptions were developed by the first author and discussed and revised with two of the co-authors.
The categories of description were further revised by presenting the analysis to external researchers, including providing example transcripts.
The two steps of revising the categories of description with internal and external researchers was done to adhere to the quality criteria proposed by Marton and Booth \cite{Marton1997}; that the categories should have a clear relation to the phenomenon under study, have a clear logical and hierarchical relationship to each other, and that they are parsimonious.
Discussions with co-authors and external colleagues was repeated until a final set of categories was agreed upon.

Phenomenography, as a qualitative research approach, still has to consider issues regarding validity and reliability, including regarding if the resulting outcome space would be significantly different with a larger study population.
In the current study, the number of participants was 18 in total, which falls within a typical accepted range for phenomenographic studies in order to capture maximum variation \cite{Kerlind2003}.
Further, in phenomenographic research, validity is often checked based on communicative and pragmatic grounds \cite{kerlind2005}.
Regarding communicative validity checks, this was checked by presenting results at various meetings with external researchers, as well as the other intended audience which are teachers.
Pragmatic validity was demonstrated through how the findings expand understanding of quantum teleportation conceptions and inform teaching of quantum physics using QPic.
Finally, reliability of the findings was ensured through dialogic reliability.
That is, the involved researchers were in full agreement of the validity of the categories and each researcher's interpretation was adhered to.

\section{Results}
\label{sec:results}
\noindent
When working with E1 and E2 during lesson two, and in answering the exam question at the end of the upper-secondary students course, the participants exhibited how varying focus on time and measurement affected their understanding of quantum teleportation as described using QPic.
In particular, we identified four categories of qualitatively different ways of understanding how a complete quantum state can be transferred across space and time using entanglement during the intervention (Table \ref{tab:categories}), i.e., a model of the principle of quantum teleportation: \textit{1. Teleportation is physical transportation of a quantum state}, \textit{2. Measurement allows for reconstitution of a quantum state}, \textit{3. Entanglement and measurement as ways of connecting particles to allow reconstitution of a quantum state}, and \textit{4. Entanglement and measurement as ways of connecting particles to allow teleportation}.
\begin{table}
\caption{\label{tab:categories}%
The outcome space with summarized descriptions for each category.
}
\begin{ruledtabular}
\begin{tabular}{ p{0.39\textwidth} p{0.58\textwidth} }
Category&
Description\\
\colrule
1. Teleportation is physical \newline transportation of a quantum state&
Quantum teleportation involves physical transportation of a quantum state from Alice to Bob. Time is not considered in diagrams and entangled particles are not present in explanations.\\

2. Measurement allows for \newline reconstitution of a quantum state&
Teleportation no longer relies on physical transportation and the need for conducting measurements on the state to be teleported is identified. `When' teleportation happens is still unclear, and the role of entanglement is not discerned. Attempts are made to connect diagrams to physical experiments.\\

3. Entanglement and measurement \newline as ways of connecting particles to \newline allow reconstitution of a \newline quantum state&
A stronger emphasis is put on the concept of entanglement, but only in the sense that Alice and Bob have access to one particle each of an entangled pair. Teleportation still relies on some reconstitution process that requires Bob to have complete knowledge of the quantum state to be teleported. Time is seen as necessary to consider in diagrams, and teleportation is identified as non-instantaneous. Attempts are made to connect diagrams to physical experiments.\\

4. Entanglement and measurement \newline as ways of connecting particles to \newline allow teleportation&
Here, entanglement is central to quantum teleportation and would not be possible without it, and all three particles are identified as becoming intimately connected through the initial entanglement and the BSM measurement. Superposition is identified to be a relevant concept. Relativity is identified as not violated due to the need for classical communication. Diagrams are used to connect abstract concepts to physical experiments.\\
\end{tabular}
\end{ruledtabular}
\end{table}
Reconstitution is used to denote that Bob can somehow change their state ($B$) into the state to be teleported ($A'$), given that Bob receives complete knowledge about $A'$.

Here, we present the outcome space as descriptions with example excerpts of each of the identified categories 1 through 4, where Category 1 represents the hierarchically lowest category, and Category 4 the highest.

\subsubsection{Teleportation is physical transportation of a quantum state}
\noindent
In this category, answers to problems and peer discussions make use of concepts, ideas, and tools provided by lesson one or two, as well as previous instruction during the course without being clearly related to one another or used in accordance to established definitions or rules.
This leads to solutions and discussions not making much sense from a physics point of view.
An example of this is given by the following answer, by S31, to an exam question asking students to motivate why quantum teleportation requires entangled particles:
\begin{quote}
    S31: \hspace{1em} Quantum teleportation with entangled particles means that the object never stops existing, only that it is transported through the wire.
\end{quote}
Here, the student is drawing upon concepts that had been introduced throughout the intervention, but conflates the mathematical rule of how boxes are allowed to move across wires in QPic with a physical movement of the state box in `reality.' During the group work session in lesson two, similar confusions between mathematical rules and physical explanations were evident as students were pointing and moving their fingers across the wire signaling that the state box, that is, the quantum state or particle to be teleported, was allowed to physically move as long as the correct mathematical operations were applied to the diagram.

\subsubsection{Measurement allows for reconstitution of a quantum state}
\noindent
This category signifies a meaningful shift in interpreting the diagrams the students had been introduced to.
Namely that it is acknowledged that there is a difference, although nebulous, between mathematical rules and operations with wires and boxes and the physical interpretations of these.
In using the mathematical tools of boxes and wires in this category, there is still some notions of the sense of physical transportation in explanations.
However, a meaningful difference is that it is asserted that the initial quantum state to be teleported is destroyed when applying the cap box, since it is recognized as being some kind of measurement and that measurements disturb, or destroys, quantum states.
When working on E1, S24 and S30 comes to the following conclusion:

\begin{quote}
    S30: \hspace{1em} Ok lets go to the next part. Do we need to change anything in our solution if we know the process involves quantum particles? \newline
    S24: \hspace{1em} Hmm. \newline
    S30: \hspace{1em} I don't... really know what... \newline
    S24: \hspace{1em} Well isn't the solution the same as the one we wrote down? \newline
    S30: \hspace{1em} Um. But when we're working with quantum particles this thing [points to the cap operation] can give an output that is a bit... wrong somehow. \newline
    S24: \hspace{1em} Hmm... \newline
    S30: \hspace{1em} Maybe we just ask the teacher. \newline
    S24: \hspace{1em} Yes but, well don't we just need to add this signal thing [classical communication] so that we know how the particle interacts with the entangled one. Then... \newline
    S30: \hspace{1em} Yes! That's correct! Then we can see what happens. Let's write it down \newline
    [Both start writing] \newline
    S30: \hspace{1em} But wait, why are we doing classical communication? To ensure the quantum state is the same or what should we say [gestures at the right side of the diagram, see Fig. \ref{fig:teleportation}a]? \newline
    S24: \hspace{1em} Um, yes. Because to know how we should change the other particle to get the same state as the original one, we need it.
\end{quote}

In the excerpt, S24 directs the discussion towards the idea that applying the cap box indicates doing some kind of measurement on two particles, $A'$ and $A$.
Then, there is a need for classical communication between Alice which allows Bob to somehow get the state to be teleported.
Here the students try to connect the diagrammatic solution to some physical reality, but without discerning the role of entanglement.
Later in their discussion, S24 and S30 concludes that the classical communication lets Bob know everything about the original state of $A'$ with no mention of entanglement, but that the wires in the diagram connects $A'$ to Bob.

As such, although they try to connect the diagrammatic solution to some physical reality, the students have not discerned the role of entanglement and instead seem to infer that it is possible, using a cap test, to somehow determine all information necessary to fully reconstruct the quantum state at another place.

\subsubsection{Entanglement and measurement as ways of connecting particles to allow reconstitution of a quantum state}
\noindent
In categories 3 and 4, entanglement plays more of a central role, as evident by the following exam question (same question as described in category 1) answer provided by S29:
\begin{quote}
    S29: \hspace{1em} That particles are entangled means that their properties somehow reflects, or are `dependent' on each other.
    When one is observed you instantly know something about the corresponding property of the other particle.
    It is this fact that is necessary for quantum teleportation to be possible.
\end{quote}
Further, when describing the quantum teleportation protocol, S29 states that Bob can somehow manipulate particle $B$ to become a copy of $A'$ since the entanglement between $A$ and $B$, along with classical communication about the output of Alice's BSM, provide complete information about $A'$.
From the written explanation, it is not discernible whether S29 considers `complete information about $A'$' to be connected to a classical understanding properties and observations.

What more clearly distinguishes Category 3 from 4, however, is the idea that teleportation still requires complete knowledge about $A'$, and that this is somehow part of the information Alice sends.
The following segment by students S14, S15, and S34 when working on E2, illustrates this way of describing teleportation, where entanglement is recognized as needed for teleportation, but described more in terms of being used as a key to decrypt Alice's classically communicated information.
\begin{quote}
    S34: \hspace{1em} Ok, so this [points to Fig. \ref{fig:exercises}b] is just teleportation. But, I'm not sure why the initial state comes from here [gestures to the text] and not from here [circles the EPR source]? \par
    S15: \hspace{1em} But if you just consider it as equal to this [shows their drawing of a diagram with boxes and wires similar to Fig. \ref{fig:teleportation}a]? \par
    S34: \hspace{1em} Ah, yes of course. You are correct. And then we have the classical communication. It's just teleportation of quantum particles. But then, does the teleportation happen faster or slower than the speed of light? \par
    S15: \hspace{1em} Well, it must be slower \par
    S14: \hspace{1em} Yes... \par
    S34: \hspace{1em} Why? Because \par
    S15: \hspace{1em} Think if you're Bob, and looks at Alice... I mean you can never observe the outcome of their results faster than the speed of light, even if something is happening with the two entangled particles instantaneously. \par
    S14: \hspace{1em} Yes, precicely! \par
    S34: \hspace{1em} Ah, ok. That sounds like a good thought. The best idea we have. Plus that Bob then gets all information through the outcome of the measurement and his entangled particle. \par
    S14: \hspace{1em} Perfect, and the entangled particle we need because.... \par
    S15: \hspace{1em} Well you know... Since Bob can use it to know, using Alice's information, the initial state [$|\psi\rangle_{A'}$] and then copy it.
\end{quote}

\subsubsection{Entanglement and measurement as ways of connecting particles to allow teleportation}
\noindent
Category 4 was rarely seen in the group work session for the upper-secondary school students.
Rather, it emerged most clearly when the pre-service teachers were working with E2:
\begin{quote}
    S3: \hspace{1em} So I think it's like this. The initial state is what we want to measure so that it can be teleported. And the teleported state is after measurement and that the entangled particle Bob has from the entangled pair is manipulated so that it becomes equal to the initial state. But we still don't know everything about the initial state. But then Bob can measure the teleported state to investigate what the initial state was. \par
    S2: \hspace{1em} Yeah, that's good. But we also need to connect this [gestures to the picture], to the wires. Like what does a wire actually mean? \par
    S1: \hspace{1em} Well, let's just do it and draw. \par
    S2: \hspace{1em} So then it's like the same as before [points to E1]? \par
    S3: \hspace{1em} I did it like this [shows their diagram to the others]. \par
    S1: \hspace{1em} Then we think like something is happening here, a measurement [points to the cap test]. \par
    S2: \hspace{1em} Yeah, and then a correction box at Bob to show how to manipulate his entangled particle so that it actually is the same as the initial state. \par
    S3: \hspace{1em} Mm. Because when we do this [points to BSM in Fig. \ref{fig:exercises}b] it really connects the initial state to Bob through entanglement [points to the `cup' state].
\end{quote}
During this episode, S3 identifies that what happens when Alice does her measurement is that all particles becomes connected through Alice's BSM and entanglement between $A$ and $B$.
As such, this category differs from the previous in that it is recognized that the measurement Alice does connects the initial state to Bobs part of the entangled pair somehow.
Additionally, S3 communicates that neither Alice nor Bob knows the full state of the teleported one, merely that entanglement and the BSM makes it so that with some manipulation by Bob they can be sure that the teleported state is in fact the same as the initial state.

Here, we see an idea about teleportation communicated that is akin to the description provided in Sec. \ref{sec:connecting_teleportation}.
Specifically that the teleportation protocol involves Bob manipulating $B$ based on the classical information provided by Alice.
Thus, in this category copying \footnote{which is of course not possible as per the no-cloning theorem \cite{Wootters1982}.} and reconstitution of a specific state based on having complete knowledge about the initial state $|\psi\rangle_{A'}$---something we see as still present in Category 2 and 3---is impossible.

Finally, when working with E2, S29 brought up superposition, drawing upon previous knowledge about the concept as taught prior to the intervention in connection to mechanical waves and an introduction to QP, to argue that complete information about the state of $A'$ is not possible to fully extract:
\begin{quote}
    S29: \hspace{1em} Well, I mean Bob can't manipulate his state until Alice sends the outcome of the measurement, which says something about the quantum state to be teleported's superposition. When receiving information about the measurement from Alice, Bob can then know what needs to be done to his particle in order to make it so that his particle's superposition is the same as the initial state. 
\end{quote}

\section{Discussion}
\label{sec:discussion}
\noindent
The four identified categories of description presented in this paper (Tab. \ref{tab:categories}) provides several implications for teaching and learning QP at the pre-university level, including affordances and limitations of using QPic employed in the current study, and the role of various concepts for students understanding of quantum teleportation.
Here, we first discuss specific concepts that emerged as central differentiators between categories, which expands upon the answer to RQ1, which is primarily answers in the results and Tab. \ref{tab:categories}.
Then, the results are discussed in relation to the role of mathematical formalism in conceptual understanding, that serves to shed more light on RQ2.
Finally implications for teaching and learning QP are discussed in relation to the current study.

\subsection{Key conceptual hurdles for understanding quantum teleportation}
\noindent
The categories of description revealed different conceptions of measurement in QP.
Category 1 stood out by being the one with a strong classical notion of measurement as something passive that does not have any effect on the process or system being observed.
In Category 2, we see a transition towards measurement being something that disturbs a quantum state, but still something that allows one to gain complete information about a quantum state after measurement.
Then, it is still possible to conduct measurements on a quantum state and fully determine everything about it, which still is within the classical realm of conceiving measurement.
The more sophisticated, disciplinary accepted description of measurement comes to the fore in categories 3 and 4, where Category 4 fully incorporates the idea that the BSM on $A'$ and $A$ is a way to couple $A'$ to $B$ via the entanglement between $A$ and $B$.
In specifically Category 4, we thus find a view of measurement to be an active process, and something that determines rather than revealing outcomes, thus fundamentally affecting the system that is being measured.

Regarding how time is conceived or being aware of, a noticeable difference can be identified if grouping together categories 1 and 2, and 3 and 4 respectively.
In the first two categories, where the process of `teleportation' includes some physical transportation or reconstitution of a quantum particle, there seem not to be a direction of time in the diagrams per se.
Additionally, in these categories the classical notion of transportation and communication is at the core of teleportation, leading to conceptions of quantum teleportation where entanglement is not a relevant concept to consider.
In contrast, the temporal recognition in the diagrams is more clearly connected to a physical experimental situation, leading to conceptions of quantum teleportation akin to categories 3 and 4 having to become aware of entanglement.
The struggle of successfully recognizing and dealing with a temporal dimension in diagrams may connect to previously identified struggles involving time and QP, such as inferring temporal directions in quantum atomic models, diagrams involving the wave equation, and general problems involving time evolution in QP \cite{Passante2019, Krijtenburg-Lewerissa2017, Singh2015}.
The identified struggles of dealing with temporality may connect to complications involving time and QP, such as the indeterministic nature of measurement outcomes on systems in superposition and that correlations due to entanglement seemingly introducing faster-than-light (FTL) communication, both conceptual challenges requiring sophisticated temporal reasoning.

Next up is the role of entanglement, and here we can focus on categories 3 and 4, as the concept was seemingly absent in the lower categories.
The differences in describing entanglement in Category 3 versus 4 boils down to the fact that in Category 4 it is identified that the system of particles is understood to be in a state similar to Eq. \ref{eq:before_measurement}, and that Alice only needs to convey classically the results from their BSM. 
In contrast, Category 3 exhibit a more fuzzy view of why exactly entanglement is necessary, focusing on the assumption that all information about the state to be teleported, $A'$, is communicated to Bob via a combination of Alice communicating results from their BSM and the fact that Bob has access to one of the originally maximally entangled particles, $B$.
While this description may sound expert-like (see, for example, the similarity to \cite{Bennett1993}), a crucial difference lies in the fact that in Category 3 there persists an assumption that in order to reconstruct the state $|\psi\rangle_{A'}$, Bob needs complete knowledge about $A'$.

Previous research has identified the challenge of teaching the concept of entanglement, in part due to it being closely connected to philosophical foundations of QP, as well as it seemingly allowing for FTL communication or involving some kind of physical connections \cite{Singh2008, Mller2002}.
From our results, it seems as though QPic in itself does little to address these previously identified struggles.
However, the intervention in this study did not focus much on connecting the diagrammatic representation of entanglement (the cup-state) to physical experimental situations beyond stating that the quantum particle $A'$ Alice wants to send to Bob could be in a superposition of states similar to Eq. \ref{eq:superposition}, and a brief discussion about measurements are considered to be an active process in QP (as opposed to being passive, non-affecting, in classical physics).
This, coupled with the relatively complex conception of the role of entanglement in quantum teleportation identified in categories 3 and 4, may hint that QPic, together with purposeful teaching, could be a fruitful mathematical tool to use as a basis for teaching QP. 

Finally, the categories of description revealed different conceptions of measurement in QP.
Category 1 stood out by being the one with a strong classical notion of measurement as something passive that does not have any effect on the process och system being observed.
In Category 2, we see a transition towards measurement being something that disturbs a quantum state, but still something that allows one to gain complete information about a quantum state after measurement.
Then, it is still possible to conduct measurements on a quantum state and fully determine everything about it, which still is within the classical realm of conceiving measurement.
The more sophisticated, disciplinary accepted description of measurement comes to the fore in categories 3 and 4, where Category 4 fully incorporates the idea that the BSM on $A'$ and $A$ is a way to couple $A'$ to $B$ via the entanglement between $A$ and $B$.
In specifically Category 4, we thus find a view of measurement to be an active process, and something that determines rather than revealing outcomes, thus fundamentally affecting the system that is being measured.

\subsection{The role of mathematical formalism in conceptual understanding}
\noindent
The mathematical formalism used to teach quantum teleportation (QPic) in the current study has been addressed as a `simplified' diagrammatic approach, which does not mean that it is `simplistic,' or `dumbed down.'
However, as is identified in our results, specifically in Category 1, there is still meaningful work to be done in order for students to make disciplinary accepted connections between a mathematical, diagrammatic solution to a problem and the corresponding physical phenomenon to be described by said solution.
Conceptions of quantum teleportation in Category 1 primarily, and to some extent Category 2 is akin to how students struggle with conceptualizing models as a simplified, or ideal version of the world \cite{KildeLfgren2023}, and how the same mathematical construct can have widely different interpretations depending on what meaning is given to each symbol in an equation (or type of wire, block, and rule in QPic) \cite{Redish2015}.
As such, the mathematical struggles identified in these lower categories provide essential further insight into previous work on teaching QP using the same diagrammatic approach \cite{Dndar-Coecke2023, Dndar-Coecke2025}.
Namely that one main argument used has been that QPic is more intuitive than traditional mathematical formalisms used to teach QP, in the sense that students are able to make use of it after rather little instruction.
Our findings problematize this notion in that the problem of connecting mathematical solutions and descriptions to physical phenomenon may persist even if the mathematics are more approachable.

\subsection{Implications for teaching and learning}
\noindent
As has been discussed previously, the traditional historical or semi-historical approach as a way to introduce QP in upper-secondary school physics classrooms may not communicate adequately the modern state of QP knowledge and utility \cite{Stadermann2019, Michelini2021}.
Our results indicate that teaching quantum phenomena such as quantum teleportation using a simplified diagrammatic approach may be a fruitful way to introduce new ways of thinking about connections between mathematical formalisms and the quantum world.
However, this study also provides further insight into students' ways of understanding quantum teleportation.
This includes making educators aware of the role or temporality, entanglement, and measurement when students try to make sense of quantum teleportation.
Specifically, educators can use our findings to design instruction that preemptively addresses the conceptual difficulties characteristic of categories 1 and 2, supporting students' progression toward more sophisticated ways of understanding quantum teleportation.

As a final note, the European Competence Framework for Quantum Technologies \cite{EuropeanCommisisionDirectorate-GeneralforCommunicationsNetworks2024} identifies the need for differentiated expertise in the quantum industry.
The diagrammatic approach used as the context for the current study may be particularly well-suited for further investigation to be used to introduce students to modern aspects of QP, since it builds the foundation for later studying ZX-calculus as a way to understand quantum theory \cite{Coecke2017}.
For teacher education, it may prove particularly useful as it opens up avenues to start from modern quantum experiments and applications, thus extending the future repertoire of available illustrations and experiments the can discuss and explore with students.

\section{Conclusion}
\label{sec:conclusion}
\noindent
In this paper, we investigated how upper-secondary physics students and pre-service teachers understood the principle of quantum teleportation when taught using QPic.
Through phenomenographic analysis, we identified four qualitatively different ways students experience this phenomenon (Tab. \ref{tab:categories}), thereby answering RQ1.
The categories form a hierarchical progression, particularly distinguished by varying conceptualizations of temporality in quantum processes, the role of entanglement, the act of measurement on quantum systems, and how QPic is identified to be a model of the quantum teleportation protocol.
The hierarchical structure of the outcome space indicates that Category 4 is the most complex way of understanding a diagrammatic representation of quantum teleportation identified in this study.

Regarding our RQ2 our results reveal both affordances and challenges.
The formalism provided an accessible entry point for studying quantum teleportation without requiring particular mathematical prerequisites.
However, as evidenced by the lower categories, students may make overly literal connections between mathematical operations in diagrams and the physical phenomena being described, leading to problematic conceptions about entanglement, superposition, and teleportation similar to those reported previously in the literature.

When designing teaching and learning sequences using a diagrammatic formalism similar to that employed in this study, several pedagogical implications merit consideration.
First, it is essential to explicitly address the relationship between mathematical operations and symbols in the formalism and the physical processes they represent.
Second, the temporal aspect of teleportation requires explicit attention, as recognizing the flow and direction of time in diagrams appears critical for avoiding classical transportation misconceptions.
Third, instruction should purposefully develop understanding of entanglement in relation to correlations and measurement outcomes on quantum systems.

Future studies exploring alternative approaches to teaching modern quantum physics concepts should consider not only whether the approach can be implemented at the upper-secondary school level, but also what new ways of conceptualizing quantum phenomena these approaches may introduce.
Additionally, given the prevalent issue that teachers possess limited knowledge of illustrative examples and experimental setups for modern quantum physics, alternative teaching approaches must consider implications for future teacher training.
Thus, studying not only student conceptions but also teacher and pre-service teacher conceptions becomes important, as teacher understanding significantly influences instructional design and implementation, ultimately shaping how students are introduced to various facets of modern quantum physics.

\section*{List of abbreviations}
\noindent
QP - Quantum physics \\
QPic - Simplified diagrammatic formalism \\
PER - Physics Education Research \\
E1 - Exercise 1 \\
E2 - Exercise 2 \\
BSM - Bell state measurement

\section*{Declarations}

\subsection*{Availability of data and materials}
\noindent
The data that support the results in this study are not publicly available as to preserve the privacy and anonymity of participants but are available from S.K-W. and J.E. upon reasonable request.

\subsection*{Competing interests}
\noindent
S.K-W., A.J., and J.E. declare no competing interests.
A.P. is employed by Quantinuum, who is the publisher of the work Quantum In Pictures which was used to develop the intervention in this study.

\subsection*{Funding}
\noindent
S.K-W. acknowledge the support from The Royal Society of Arts and Sciences in Gothenburg.

\subsection*{Authors' contribution}
\noindent
\textbf{S.K-W.}: Conceptualization, Data curation, Formal analysis, Investigation, Methodology, Project administration, Resources, Validation, Visualization, Writing -- original draft, Writing -- review \& editing. \\
\textbf{A.J.}: Formal analysis (supporting), Methodology (supporting), Validation, Visualization (supporting), Writing -- review \& editing \\
\textbf{A.P.}: Resources, Validation, Writing -- review \& editing \\
\textbf{J.E.}: Data curation, Formal analysis (supporting), Investigation (supporting), Methodology (supporting), Project administration, Resources, Supervision, Validation, Writing -- review \& editing.

\subsection*{Acknowledgements}
\noindent
S.K-W. and A.P. would like to thank Muhammad Hamza Waseem and Caterina Puca for valuable discussions about experiences on teaching the contents of Quantum in Pictures, and for commenting on a previous version of the paper.
S.K-W. and A.P. are also grateful to Fredrik Lannestam Holmelin for connecting them, thus making this collaboration possible.
The educational material presented in Sec. \ref{sec:context} and used as a basis for the intervention in this study was developed based on the work “Quantum In Pictures” by Professor Bob Coecke and Dr Stefano Gogioso, © 2022 Quantinuum Ltd.

\bibliography{refs.bib}

\end{document}